\documentstyle[seceq,epsf]{ptptex}
\title{Single-pole Nature of the Detectable $\Lambda(1405)$}

\author{Khin Swe {\sc Myint}$^{*}$, Yoshinori {\sc Akaishi}$^{**}$, 
    Maryam {\sc Hassanvand}$^{***}$ and Toshimitsu {\sc Yamazaki}$^{**,****}$}

\inst{$^{*}$Department of Physics, Mandalay University, Mandalay, Union of Myanmar \\
$^{**}$RIKEN Nishina Center, Wako, Saitama 351-0198, Japan \\
$^{***}$Department of Physics, Isfahan University of Technology, Isfahan 84156-83111, Iran \\
$^{****}$Department of Physics, University of Tokyo, Tokyo 113-0033, Japan }

\recdate{%      %Editorial Office will fill in this.
}

\abst{We have investigated the single-pole nature of the detectable $\Lambda(1405)$ in detail, by employing a chiral model having double poles of a $\bar KN$-$\pi \Sigma$ coupled channel $T$-matrix. The effects of the 1st pole and the 2nd pole on the observed $\pi \Sigma$ mass spectrum in the $\Lambda(1405)$ region were analyzed almost separately by means of the Generalized Optical Potential. It is concluded that the 1st pole is responsible for both peak structures seen in the $T_{\pi \Sigma \leftarrow \bar KN}$($T_{21}$) and $T_{\pi \Sigma \leftarrow \pi \Sigma}$($T_{22}$) invariant-mass spectra, and the 2nd pole due to an energy-dependent chiral interaction provides continuum amplitudes affecting the shape of the peak structures of the 1st pole in experimentally observable spectrum of the $\Lambda(1405)$.}

\begin{document}

\maketitle

\section{Introduction} 

The $\Lambda(1405)$ is nominally accepted as being a quasi-bound state of the $\bar KN$ system \cite{Dalitz59} decaying to $\pi \Sigma$, or/and a resonance state in the $\pi \Sigma$ channel coupled to a $\bar KN$ state (Feshbach resonance \cite{Feshbach58}), with strangeness $S$=$-1$ and isospin $I$=0. The precise mass and width of the  $\Lambda(1405)$ is still somewhat ambiguous due to not only limited experimental data, but also theoretical problems of model-dependent analyses, one of which is a single- or double-pole nature of the $\Lambda(1405)$, 
as discussed in this paper. 
The phenomenological analyses \cite{Dalitz91,Esmaili10,Hassanvand13} contributing to the present PDG value, a mass of $1405.1^{+1.3}_{-1.0}$ MeV$/c^2$ and a width of $50.5 \pm 2.0$ MeV \cite{Patrignani16}, were conducted on the basis of the {\it single-pole nature} of the $\Lambda(1405)$. 

Recently, a very large number of data sets on the photo-production of the $\Lambda(1405)$ were provided in the $\gamma p \rightarrow K^+\pi^{0 \pm}\Sigma^{0\mp}$ reaction experiment at center-of-mass energies of $1.95 < W < 2.85$ GeV by the CLAS collaboration \cite{Moriya13}. There have appeared theoretical papers \cite{Roca13,Mai15,Nakamura14} analyzing the CLAS data based on chiral dynamics, where the {\it double-pole structure} \cite{Oller01,Jido03} of the $\Lambda(1405)$ is strongly claimed.  As summarized in Meissner and Hyodo's report \cite{Meissner15}, the 1st pole coupled mainly to the $\bar KN$ channel lies at around $1421 \sim 1434$ MeV$/c^2$ mass region, fairly higher than the PDG value, though the 2nd pole coupled to the $\pi \Sigma$ channel distributes in a rather wide mass-range. A recent phenomenological analysis of the CLAS data based on the single-pole nature has extracted pole positions with statistical confidence \cite{Hassanvand17} consistent with the PDG value of the $\Lambda(1405)$. 

Thus, "whether the $\Lambda(1405)$ is of a single pole or of double poles"\cite{Akaishi10} is a current issue to be studied as a basic problem for practical data analyses. Since $\Lambda(1405)$ is observed mainly through the $\pi \Sigma$ mass spectra, we introduce the concept of a "detectable $\Lambda(1405)$" as a pole or poles of a $\bar KN$-$\pi \Sigma$ coupled-channel $T$-matrix, which make(s) peak structures in the $\pi \Sigma$ invariant-mass spectra to appear below the $\bar KN$ threshold. We employ the $I$=0 part of a two-channel treatment of the chiral dynamics presented by Hyodo and Weise \cite{Hyodo08}, and investigate the issue by asking how their double poles relate to the formation of peak structures in the $\pi \Sigma$ mass spectra. For this purpose, we use generalized optical potentials. A main aim of this paper is to derive theoretically the nature of the detectable $\Lambda(1405)$ through its formation mechanism. 

This paper is organized as follows. In Section 2 we give a formulation by developing the use of generalized optical potentials. In Section 3 the formation mechanism of peak structures in the $\pi \Sigma$ mass spectra is investigated in relation to the chiral double poles. The single-pole or double-pole nature of the $\Lambda(1405)$ is discussed in Section 4. Conclusions are given in Section 5.

\section{Formulation}

\subsection{Problem and model setting}

\begin{figure}[hbtp]
\begin{center}
\hspace{0.5cm}
\vspace{-0.5cm}
\epsfxsize=8cm
\epsfbox{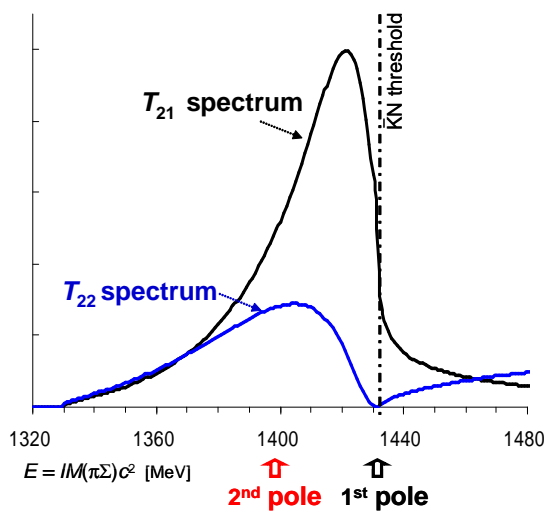}
\vspace{0.5cm}
\end{center}
\caption{$I$=0 $\pi \Sigma$ invariant-mass spectra, $T_{21}$ spectrum and $T_{22}$ spectrum, obtained from the two-channel treatment of the chiral dynamics of H-W\cite{Hyodo08}. The double-pole positions are also shown.}
\label{fig1}
\end{figure}

As stated in Introduction, we define the detectable $\Lambda (1405)$ as a pole or poles of an $I$=0 $\bar KN$-$\pi \Sigma$ coupled-channel $T$-matrix that make(s) peak structures in the detectable $\pi \Sigma$ mass spectra below the $\bar KN$ threshold. In order to make our logic transparent we set a simplified model without any loss of essential dynamics: we employ a two-channel treatment of the $I$=0 part of the chiral dynamics given by Hyodo and Weise (H-W) \cite{Hyodo08}, 
\begin{equation}
V_{ij}(\sqrt s)=-\frac{C_{ij}}{4 f^2} (2\sqrt s -M_i-M_j) \sqrt{\frac{E_i+M_i}{2M_i}} \sqrt{\frac{E_j+M_j}{2M_j}}, 
\label{chiral}
\end{equation}
which is the leading-order Weinberg-Tomozawa term of H-W's eq.(2), with the loop function, $G_i$ of H-W's eq.(3), 
\begin{equation}
G_i(\sqrt s)=\frac{2M_i}{(4 \pi)^2} \left\{ a_i(\mu) + {\rm ln} \frac{M_i^2}{\mu^2} + \frac{m_i^2-M_i^2+s}{2s}{\rm ln} \frac{m_i^2}{M_i^2} + 
\frac{\bar q_i}{\sqrt{s}}{\rm ln} \frac{\phi_{++}(s)\phi_{+-}(s)}{\phi_{-+}(s)\phi_{--}(s)} \right\}. 
\label{loop}
\end{equation}
The double poles exist at $\Xi_{\rm pole}=M_{\rm pole}c^2-i\Gamma_{\rm pole}/2$ as 
\begin{eqnarray}
{\rm 1st~pole :}~~~~~~~\Xi_{\rm 1^{st}pole}=1432-i17~{\rm MeV},  \\ 
{\rm 2nd~pole :}~~~~~~\Xi_{\rm 2^{nd}pole}=1398-i73~{\rm MeV}. 
\label{poles} 
\end{eqnarray}
In the present model the $I$=0 $T$-matrix consists of basic $T^{I=0}_{\pi \Sigma \leftarrow \bar KN}$ and $T^{I=0}_{\pi \Sigma \leftarrow \pi \Sigma}$ matrices. Hereafter, we use simplified notations as 
\begin{equation}
T_{21} \equiv T^{I=0}_{\pi \Sigma \leftarrow \bar KN}~~~ {\rm and}~~~T_{22} \equiv T^{I=0}_{\pi \Sigma \leftarrow \pi \Sigma},
\end{equation} 
where 1 and 2 indicate the $\bar KN$ and $\pi \Sigma$ channels, respectively. The basic $\pi \Sigma$ invariant-mass spectra are 
\begin{eqnarray}
T_{21}~{\rm spectrum :}~~~~~~\vert T_{21}(E) \vert^2~ k_2(E),  \\ 
T_{22}~{\rm spectrum :}~~~~~~\vert T_{22}(E) \vert^2~ k_2(E),
\label{spDef} 
\end{eqnarray}
where $E$ is the total energy ($\sqrt s$) and $k_2$ is the $\pi$(or $\Sigma$) momentum in the center-of-mass system of on-shell decay particles, $\pi \Sigma$. The two basic spectra of the present model are shown in Fig. \ref{fig1} together with the double-pole positions.
%
%%%%% Fig.1 %%%%%
The peak position of the $T_{22}$ spectrum is close to the position of the 2nd pole. This proximity could induce a misunderstanding about the relation between the 2nd pole and the peak structure of the $T_{22}$ spectrum. The relation is a main problem to be solved in this paper. 

\subsection{Generalized optical potential}

In order to investigate the property of double poles appearing in $T$-matrices of the present chiral dynamics we introduce generalized optical potentials for two coupled channels of $\bar KN$ (channel 1) and $\pi \Sigma$ (channel 2). The fundamental quantities that are relevant to the $\pi \Sigma$ invariant-mass spectra are the transition-matrix elements, $T_{21}$ and $T_{22}$, which obey the following coupled-channel $T$-matrix equation: 
\begin{equation}
\left[ \begin{array}{cc} T_{11}& T_{12} \\
T_{21}& T_{22} \end{array} \right] = \left[
\begin{array}{cc} V_{11}& V_{12} \\
V_{21}& V_{22} \end{array} \right] + \left[
\begin{array}{cc} V_{11}& V_{12} \\
V_{21}& V_{22} \end{array} \right] \left[
\begin{array}{cc} G_1& 0 \\
0& G_2 \end{array} \right] \left[
\begin{array}{cc} T_{11}& T_{12} \\
T_{21}& T_{22} \end{array} \right], 
\label{mateq}
\end{equation}
where $G_i$ is the $i$-channel loop integral.

By the use of generalized optical potentials, defined as
\begin{eqnarray}
V^{\rm opt}_{11}(\Xi)&=&V_{11}+V_{12} \frac{G_2(\Xi)}{1-V_{22}G_2(\Xi)} V_{21},
\label{opt1}  \\
V^{\rm opt}_{12}(\Xi)&=&V_{12} \frac{1}{1-G_2(\Xi)V_{22}}, \\
V^{\rm opt}_{21}(\Xi)&=&V_{21} \frac{1}{1-G_1(\Xi)V_{11}}, 
\label{opt3} \\
V^{\rm opt}_{22}(\Xi)&=&V_{22}+V_{21} \frac{G_1(\Xi)}{1-V_{11}G_1(\Xi)} V_{12},
\label{opt4}
\end{eqnarray} 
the solutions of each $T$-matrix element are exactly given by
\begin{eqnarray}
T_{11}(\Xi)=\frac{1}{1-V^{\rm opt}_{11}(\Xi)G_1(\Xi)} V^{\rm opt}_{11}(\Xi), 
\label{sol1} \\
T_{12}(\Xi)=\frac{1}{1-V^{\rm opt}_{11}(\Xi)G_1(\Xi)} V^{\rm opt}_{12}(\Xi), \\
T_{21}(\Xi)=\frac{1}{1-V^{\rm opt}_{22}(\Xi)G_2(\Xi)} V^{\rm opt}_{21}(\Xi), 
\label{sol3}\\
T_{22}(\Xi)=\frac{1}{1-V^{\rm opt}_{22}(\Xi)G_2(\Xi)} V^{\rm opt}_{22}(\Xi), 
\label{sol4}
\end{eqnarray} 
where $\Xi$ is a complex total-energy variable in the center-of-mass system. The relation 
\begin{equation}
\{1-V^{\rm opt}_{22}(\Xi)G_2(\Xi)\}\{1-G_1(\Xi)V_{11}(\Xi)\}=\{1-V_{22}(\Xi)G_2(\Xi)\}\{1-G_1(\Xi)V^{\rm opt}_{11}(\Xi)\} 
\label{optrel}
\end{equation}
is useful for forthcoming discussions. In the case of an energy-dependent interaction the $V_{ij}$'s also depend on $\Xi$. An advantage of using the generalized optical potentials is that any effects of double poles on detectable mass spectra can be examined almost separately. 

The following relations are derived with Eqs.(\ref{opt1}) $\sim$ (\ref{sol4}) at real total energies, $E$, in the mass region below the $\bar KN$ threshold: 
\begin{eqnarray}
-\frac{1}{\pi}~{\rm Im}~T_{11}(E) = \vert T_{21}(E) \vert^2 ~(-\frac{1}{\pi})~{\rm Im}~G_2(E), 
\label{Spec1} \\
-\frac{1}{\pi}~{\rm Im}~T_{22}(E) = \vert T_{22}(E) \vert^2 ~(-\frac{1}{\pi})~{\rm Im}~G_2(E).  
\label{Spec2}
\end{eqnarray} 
The right-hand sides of Eqs.(\ref{Spec1}) and (\ref{Spec2}) are related to the detectable $\pi \Sigma$ invariant-mass spectra, $T_{21}$ spectrum and $T_{22}$ spectrum, respectively. The left-hand sides are strength functions, which are observable also through missing-mass spectra of such reactions as $d(K^-,n)X$.

\section{$T$-matrix poles and peak structures in the $\pi \Sigma$ mass spectra}

\subsection{Characteristics of $T_{\pi \Sigma \leftarrow \bar KN}$ ($T_{21}$) invariant-mass spectrum}

We investigate the characteristics of the $T_{21}$ spectrum in relation to the 1st pole, which is naturally considered to be the origin of the peak structure. The detectable $T_{21}$ invariant-mass spectrum is given by the following $T$-matrix of Eq.(\ref{sol3}) with a real total-energy (invariant-mass) variable, $E$: 
\begin{eqnarray}
T_{21}(E)=\frac{1}{1-V^{\rm opt}_{22}(E)G_2(E)} V_{21} \frac{1}{1-G_1(E)V_{11}(E)}  \nonumber \\
~~~=\frac{1}{1-V_{22}(E)G_2(E)} V_{21} \frac{1}{1-G_1(E)V^{\rm opt}_{11}(E)}, 
\label{sol3r}
\end{eqnarray}
where Eq.(\ref{optrel}) is used. We introduce an arbitrary factor, $f_{11}$, as
\begin{equation}
V_{11} \rightarrow V_{11} \cdot f_{11}  
\end{equation}
into $V^{\rm opt}_{11}$. 

\begin{figure}[hbtp]
\begin{center}
\hspace{0.5cm}
\vspace{-0.5cm}
\epsfxsize=12cm
\epsfbox{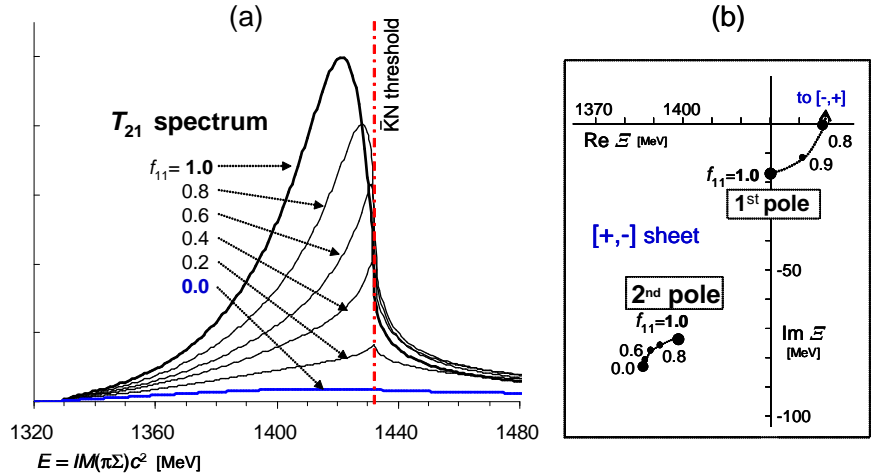}
\vspace{0.5cm}
\end{center}
\caption{(a) Change of the $T_{21}$-spectrum shape at varied $f_{11}=1 \sim 0$, where $V_{11}$ is weakened to $V_{11} \cdot f_{11}$. The $f_{11}$=1.0 (black) curve is identical to the $T_{21}$ spectrum in Fig. \ref{fig1}. ~~(b) Trajectories of the 1st and 2nd poles. The 1st pole moves towards the higher-mass side with decreasing $f_{11}$ and goes out of the $[+({\rm ch.}1),-({\rm ch.}2)]$ Riemann sheet, while the 2nd pole stays on the $[+,-]$ sheet.}
\label{fig2}
\end{figure}

Figure \ref{fig2}(a) shows the spectra at varied $f_{11}$ from 1 to 0 in steps of 0.2 together with pole trajectoris of Fig. \ref{fig2}(b). The peak structure of the $T_{21}$ spectrum at $f_{11}$=1 changes to threshold-cusp structures at $f_{11} < 0.7$, where the $\bar KN$ quasi-bound state disappears. This clearly indicates that the origin of the peak structure in the $T_{21}$ invariant-mass spectrum is the 1st pole, which satisfies $1-V^{\rm opt}_{11}(\Xi_{\rm 1^{st}pole})G_1(\Xi_{\rm 1^{st}pole})=0$. In fact, the curve at $f_{11}$=0 denotes that the 2nd pole from $V_{22}$ cannot produce any peak structure without the 1st pole. Thus, the peak structure of the $T_{21}$ spectrum is confirmed to be a typical Feshbach resonance\cite{Feshbach58} caused by the 1st pole. The shift of the peak position from the 1st pole position is due to the $\bar KN$ threshold effect, as is well-known.  
%
%%%%% Fig.2 %%%%%

\subsection{Characteristics of $T_{\pi \Sigma \leftarrow \pi \Sigma}$ ($T_{22}$) invariant-mass spectrum}

This subsection deals with an investigation into characteristics of the $T_{22}$ spectrum, which reveals the nature and origin of the $\Lambda(1405)$ from another side than $T_{21}$. The detectable $T_{22}$ invariant-mass spectrum is given by the following $T$-matrix with a real total-energy (or invariant-mass) variable, $E$:
\begin{equation}
T_{22}(E)=\frac{1}{1-V^{\rm opt}_{22}(E)G_2(E)} V^{\rm opt}_{22}(E). 
\label{sol4r}
\end{equation}
In order to examine the role of the 2nd pole, which is caused by the energy-dependent $V_{22}(\Xi)$ according to chiral dynamics \cite{Hyodo08}, we introduce two arbitrary factors, $f_{22}$ and $f_{12}$, into the optical potential as 
\begin{equation}
V^{\rm opt}_{22}=V_{22} \cdot f_{22}+V_{21} \frac{G_1}{1-V_{11}G_1} V_{12} \cdot f_{12}.  
\label{opt4r}
\end{equation}
By varying the values of the factors from one to zero, we study how the 1st pole and the 2nd pole affect the $T_{22}$ spectrum.
%
%%%%% Fig.3 %%%%%
\begin{figure}[hbtp]
\begin{center}
\hspace{0.5cm}
\vspace{-0.5cm}
\epsfxsize=12cm
\epsfbox{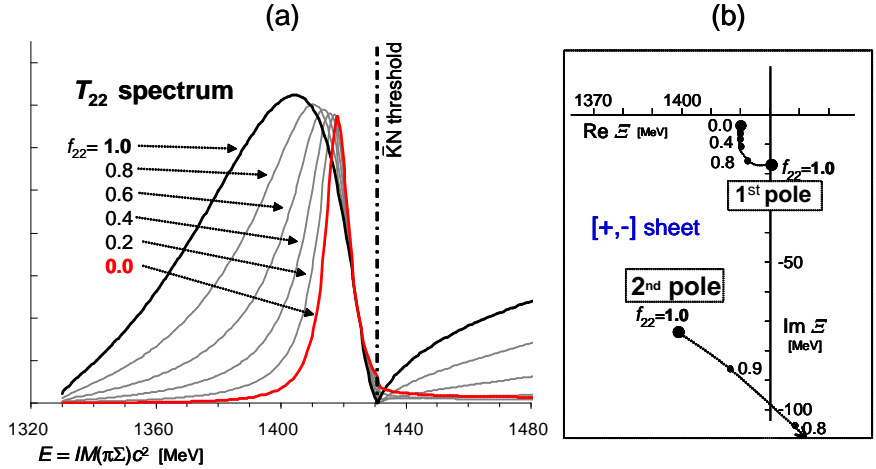}
\vspace{0.5cm}
\end{center}
\caption{(a) Change of the $T_{22}$-spectrum shape at varied $f_{22}=1 \sim 0$ with fixed $f_{12}=1$. The $f_{22}$=1.0 (black) curve is identical to the $T_{22}$ spectrum in Fig. \ref{fig1}. ~~(b) Trajectories of the 1st and 2nd poles. The 2nd pole rapidly escapes from the relevant $\Lambda(1405)$ mass region as $f_{22}$ decreases.}
\label{fig3}
\end{figure}
%

%
%%%%% Fig.4 %%%%%
\begin{figure}[hbtp]
\begin{center}
\hspace{0.5cm}
\vspace{-0.5cm}
\epsfxsize=12cm
\epsfbox{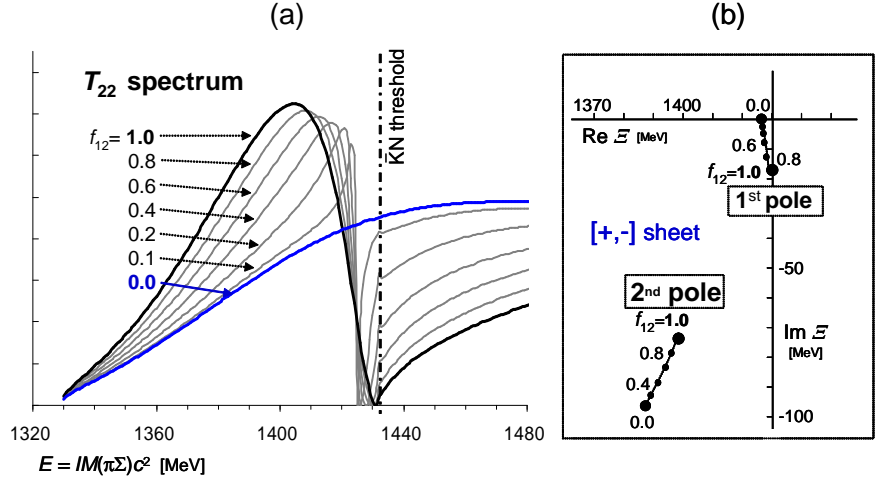}
\vspace{0.5cm}
\end{center}
\caption{(a) Change of the $T_{22}$-spectrum shape at varied $f_{12}=1 \sim 0$ with fixed $f_{22}=1$. The $f_{12}$=1.0 (black) curve is identical to the $f_{22}$=1.0 (black) curve in Fig. \ref{fig3}. ~~(b) Trajectories of the 1st and 2nd poles. When $f_{12}$ is switched on, a singular structure starts at the 1st pole position.}
\label{fig4}
\end{figure}

In the first case, $f_{22}$ is varied from 1 to 0, while $f_{12}$ is fixed at 1. The obtained spectra are shown in Fig. \ref{fig3}(a), where the spectrum retains the peak structure even at $f_{22}=0$ (red curve) when no effect is expected from the 2nd pole. This indicates that the 1st pole is the seed growing to the peak structure in the detectable $T_{22}$ spectrum ($f_{22}$=1.0 curve). As another possibility, it might be thought that the 2nd pole contribution is the difference between the $f_{22}$=1 and $f_{22}$=0 curves, which is the major part of the peak structure. However, this idea is wrong, because the 2nd pole stays far away from the relevant $\Lambda(1405)$ mass region at least for $f_{22} < 0.7$ as seen in Fig. \ref{fig3}(b). 

In the next case, $f_{22}$ is fixed at 1 and $f_{12}$ is varied from 1 to 0. Figure \ref{fig4}(a) shows the obtained spectra. It can be seen that the peak structure converges to the 1st pole position with decreasing values of $f_{12}$, and finally disappears at $f_{12}=0$. Since the interaction, $V_{22}$, is kept unchanged, the 2nd pole still exists at $f_{12}=0$ with $M_{\rm pole}=1388~{\rm MeV}/c^2,\Gamma=192~{\rm MeV}$ as denoted in \ref{fig4}(b) (see also Eq. (\ref{2ndpole})). Although one might expect a rise due to the 2nd pole at this $f_{12}=0$, the corresponding (blue) curve has no such structure below the $\bar KN$ threshold. Thus, Fig. \ref{fig4} shows again that the origin of the peak structure in the $T_{22}$ invariant-mass spectrum is not the 2nd pole.

\subsection{Moving pole for the detectable $T_{22}$ spectrum}

The next step of our investigation is to answer why the 2nd pole produces no peak structure of the $f_{12}$=0 spectrum (blue curve) in Fig. \ref{fig4}(a). The relevant 2nd pole, $\Xi_{\rm pole}$, is obtained from the following equation:
\begin{equation}
T^{-1}_{22}(\Xi_{\rm pole}) = V^{-1}_{22}(\Xi_{\rm pole})-G_2(\Xi_{\rm pole})=0,
\label{poleEq}
\end{equation}
to be 
\begin{equation}
\Xi_{\rm pole} = 1388 - i 96~~ {\rm MeV},
\label{2ndpole}
\end{equation}
as given in H-W's eq.(9) \cite{Hyodo08}, which shifts to the position of Eq. (\ref{poles}) in the $f_{12}$=1 case. 
Then, one may consider to apply a $T$ matrix with this 2nd pole, 
\begin{equation}
{\tilde T}_{22}(E) =  [~V^{-1}_{22}(\Xi_{\rm pole})-G_2(E)~]^{-1}, 
\label{T22-pole}
\end{equation}
to the $T_{22}$ spectrum at the detectable real $E$. Actually, some authors \cite{Jido03,Jido10} who claim the two pole structure of the $\Lambda(1405)$ recommend to use a Breit-Wigner amplitude, corresponding to Eq. (\ref{T22-pole}), from the 2nd pole for data analyses. But, it should be noticed that the correct $T$ matrix to be used for the observable $T_{22}$ spectrum of Eq. (\ref{spDef}) is 
\begin{equation} 
T_{22}(E)= [~V^{-1}_{22}(E)-G_2(E)~]^{-1}. 
\label{T22-detec} 
\end{equation} 

\begin{figure}[hbtp]
\begin{center}
\hspace{0.5cm}
\vspace{-0.7cm}
\epsfxsize=8cm
\epsfbox{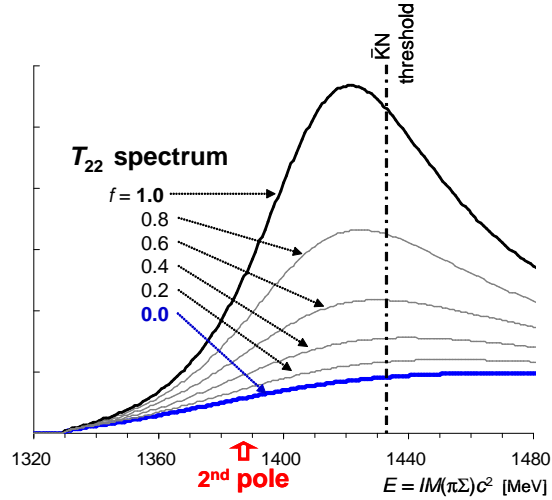}
\vspace{0.5cm}
\end{center}
\caption{Dependence of the $T_{22}$ spectra on the positive imaginary (source) term of $V_{22}$. The $f$=0.0 (blue) curve is identical to the $f_{12}$=0.0 (blue) curve in Fig. \ref{fig4}(a).}
\label{fig5}
\end{figure}

In order to reveal the difference between the effects of $V_{22}(\Xi_{\rm pole})$ and of $V_{22}(E)$, we define "reference" $T$-matrix with a parameter, $\Xi_{\rm prm}$ , to fix the $V_{22}$,
\begin{equation}
T^{\rm ref}_{22}(\Xi;\Xi_{\rm prm}) \equiv [~V^{-1}_{22}(\Xi_{\rm prm}) -G_2(\Xi)~]^{-1}, 
\label{Tref0}
\end{equation}
and investigate 
\begin{equation}
T^{\rm ref}_{22}(E;E+i~{\rm Im}\Xi_{\rm pole} \cdot f), 
\label{Tref}
\end{equation}
where $V_{22}$ includes a positive imaginary (source) term proportional to $f$. We have varied the imaginary part of $V_{22}$ with the factor $f$ ranging from 1 to 0. The resultant spectra are shown in Fig. \ref{fig5}. It is seen that, with a large value of ${\rm Im}~\Xi_{\rm pole}$ at $f$=1, a peak structure of the Breit-Wigner type appears in the spectrum. However, the peak structure decreases and finally disappears with no imaginary part at $f$=0, which is just the experimentally observable case. This is numerical evidence that the 2nd pole produces no peak in the detectable $T_{22}$ invariant-mass spectrum. 
%
%%%%% Fig.5 %%%%%
%
%
%%%%% Fig.6 %%%%%
\begin{figure}[hbtp]
\begin{center}
\hspace{0.5cm}
\vspace{-0.5cm}
\epsfxsize=10cm
\epsfbox{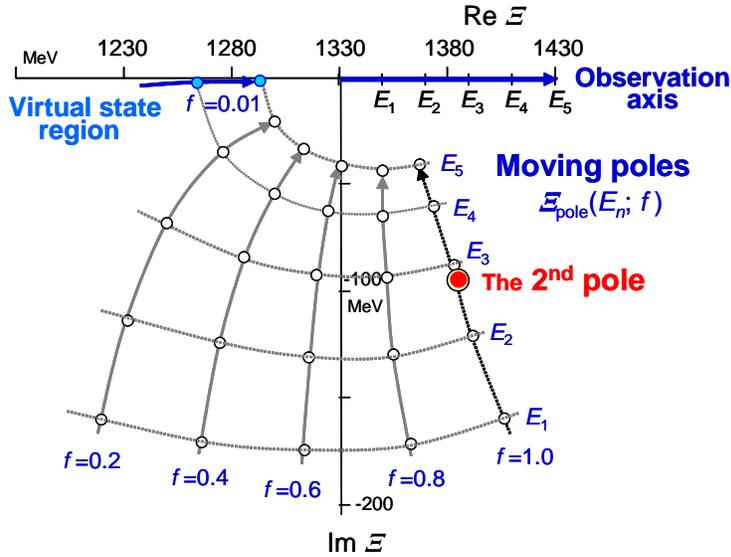}
\vspace{0.5cm}
\end{center}
\caption{Behavior of moving poles, $\Xi_{\rm pole}(E_n;f)$, defined in the text. }
\label{fig6}
\end{figure}

The pole of the reference $T$ matrix of Eq. (\ref{Tref}) is obtained as a solution of the following equation:
\begin{equation}
V^{-1}_{22}(E+i~{\rm Im}\Xi_{\rm pole} \cdot f)-G_2(\Xi)=0, 
\label{poleEq2}
\end{equation}
for given $E$ and $f$. The solution $\Xi_{\rm pole}(E;f)$ moves, through $V_{22}$ and boundary condition \cite{Akaishi08,Koike09}, as $E$ changes.  
Such a "moving pole" \cite{Koike09} is depicted in Fig. \ref{fig6} by taking discrete $E$'s from $E_1$=1350 MeV to $E_5$=1430 MeV in steps of 20 MeV. Each nearly vertical path in Fig. \ref{fig6} shows the motion of the moving pole, $\Xi_{\rm pole}(E_n;f)$, obtained from the fixed imaginary part with $f=1.0 \sim 0.01$. As the imaginary part approaches zero, the moving pole shifts in the virtual-state region apart far from the original position of the 2nd pole. Since $T_{22}(E)$=$T^{\rm ref}_{22}(E;\Xi_{\rm prm}=E~{\rm at}~f{\rm =0})$, the moving pole with $f$=0, i.e. "decaying-state pole" discussed in ref.\cite{Akaishi08,Kapur38}, is responsible for the detectable $T_{22}(E)$ of Eq. (\ref{T22-detec}). Thus, it is clearly shown that {\it the 2nd pole position of Eq. (\ref{2ndpole}) cannot be related to the peak position in the observed $T_{22}$ spectrum}. 

\subsection{Formation mechanism of the peak structure in the $T_{22}$ spectrum}

\begin{figure}[hbtp]
\begin{center}
\hspace{0.5cm}
\vspace{-0.5cm}
\epsfxsize=12.5cm
\epsfbox{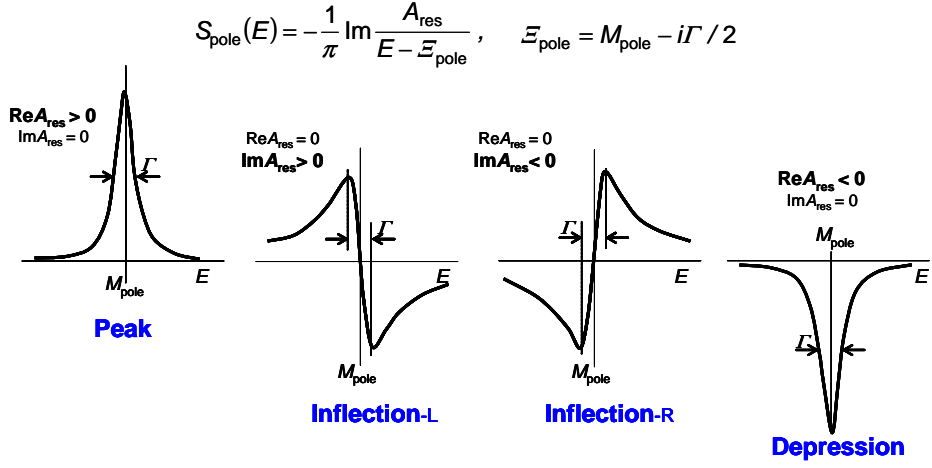}
\vspace{0.3cm}
\end{center}
\caption{Relationship between spectral shape and residue of a pole term in strength function, which was originally given in ref.\cite{Morimatsu88}. } 
\label{fig11}
\end{figure}

We want to know the formation mechanism of the peak structure in the $T_{22}$ spectrum of the coupled-channel system. Firstly we investigate the spectrum shape of the 2nd pole term. This shape problem was fully discussed in Green's function formalism by Morimatsu and Yazaki (see Fig. \ref{fig11} depicted from fig. 8 of Ref.\cite{Morimatsu88}). Following their procedure, we extract the contribution of the 2nd pole by using Eq.(\ref{Spec2}) and Eq.(\ref{sol4}) with an expansion of the denominator, $1-V^{\rm opt}_{22}(E)G_2(E)$, of $T_{22}(E)$ into a power series of $(E-\Xi_{\rm 2^{nd}pole})$, 
\begin{eqnarray}
-\frac{1}{\pi}~{\rm Im}~T_{22}(E) = -\frac{1}{\pi}~{\rm Im}~\frac{A_{\rm 2^{nd}pole}}{E-\Xi_{\rm 2^{nd}pole}} + {\rm the~remaining~part}, \\
A_{\rm 2^{nd}pole}=-V^{\rm opt}_{22}(\Xi_{\rm 2^{nd}pole})
\big/\frac{dV^{\rm opt}_{22}G_2}{d\Xi}(\Xi_{\rm 2^{nd}pole}). 
\label{opt4p2}
\end{eqnarray}
The 2nd-pole term has a spectrum shape of 
\begin{equation}
\frac{1}{\pi} \frac{({\rm Re}~A_{\rm 2^{nd}pole})\cdot \Gamma/2}{(E-E_{\rm 2^{nd}pole})^2+\Gamma^2/4}-\frac{1}{\pi} \frac{({\rm Im}~A_{\rm 2^{nd}pole})\cdot (E-E_{\rm 2^{nd}pole})}{(E-E_{\rm 2^{nd}pole})^2+\Gamma^2/4}, 
\label{poleShape}
\end{equation}
where $\Xi_{\rm 2^{nd}pole}=E_{\rm 2^{nd}pole}-i \Gamma/2$. At the pole position, $E_{\rm 2^{nd}pole}$, the $({\rm Re}~A_{\rm 2^{nd}pole})$ part gives a peak (or depression), but the $({\rm Im}~A_{\rm 2^{nd}pole})$ part exhibits an inflection\cite{Morimatsu88} of the relevant curve. 

%%%%% Fig.11 %%%%%
%
%
%%%%% Fig.7 %%%%%
\begin{figure}[hbtp]
\begin{center}
\hspace{0.5cm}
\vspace{-0.5cm}
\epsfxsize=12.5cm
\epsfbox{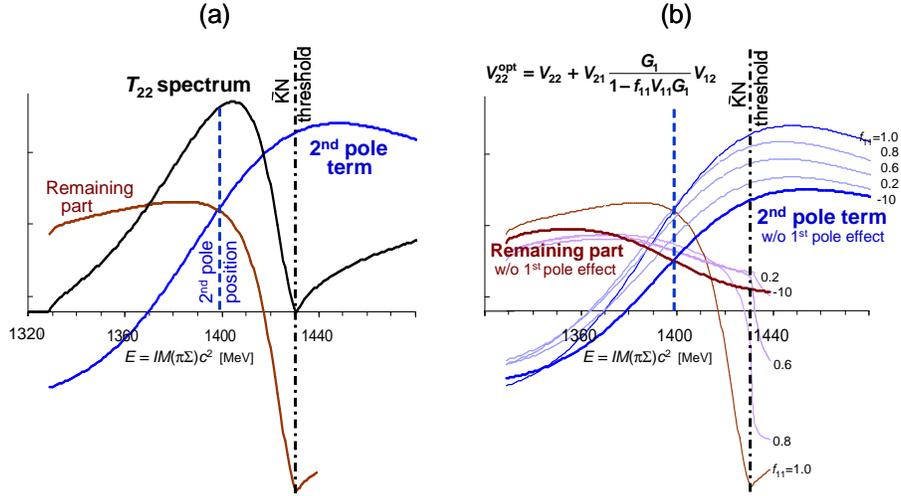}
\vspace{-0.3cm}
\end{center}
\caption{(a) Breakdown of the $T_{22}$ spectrum. The contributions of the 2nd-pole term (blue curve) and of the remaining part (brown curve) are shown. The remaining part includes the 1st-pole term. The $T_{22}$ spectrum (black curve) is identical to the $f_{22}$=1.0 (black) one in Fig. \ref{fig3}(a).~~
(b) Substantial contributions of the 2nd-pole term and of the remaining part when the 1st pole effect is removed as explained in Fig. \ref{fig8}. Both the contributions are on the same order.}
\label{fig7}
\end{figure}
%
%
%%%%% Fig.8 %%%%%
\begin{figure}[hbtp]
\begin{center}
\hspace{0.5cm}
\vspace{-0.5cm}
\epsfxsize=8.5cm
\epsfbox{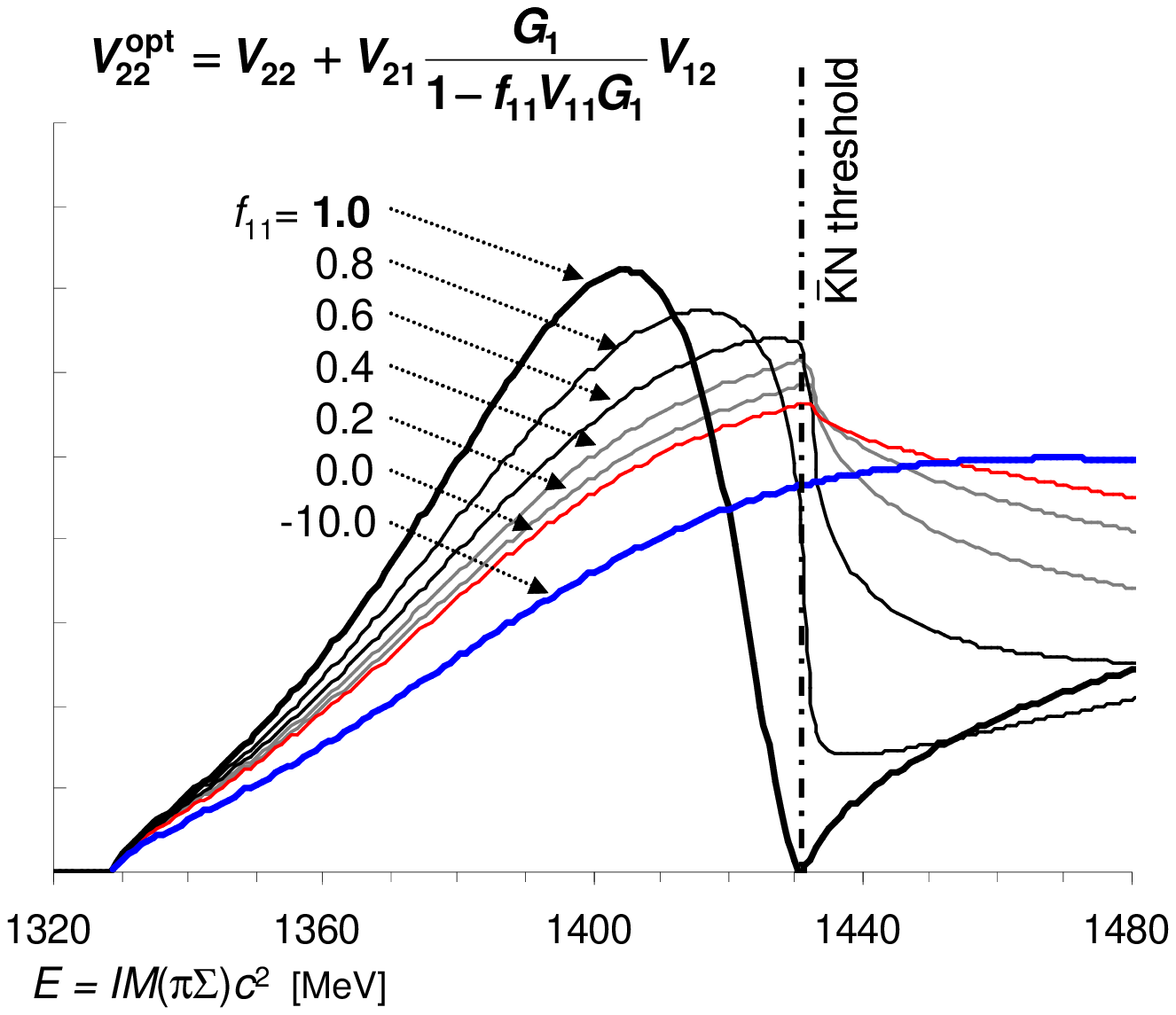}
\vspace{0.5cm}
\end{center}
\caption{(a) Effects of the position shift of the 1st pole on the $T_{22}$ spectrum. An arbitrary factor, $f_{11}$, is varied from 1 to -10, while shifting the 1st-pole position. The $f_{11}$=-10 (blue) curve is close to the $f_{12}$=0.0 (blue) one in Fig. \ref{fig4}. The $f_{11}$=1.0 (black) curve is identical to the $f_{12}$=1.0 (black) one in Fig. \ref{fig4}. (b) Movements of the 1st and 2nd poles. The 1st pole moves towards the higher-mass side with decreasing $f_{11}$ and goes out of the $[+({\rm ch.}1),-({\rm ch.}2)]$ sheet, while the 2nd pole stays on the $[+,-]$ sheet. }
\label{fig8}
\end{figure}

Figure \ref{fig7}(a) shows the 2nd-pole term (blue curve) and the remaining part (brown curve) of the $T_{22}$ spectrum (black curve). In the 2nd-pole term a plateau appears not at the pole position, but above the $\bar KN$ threshold. This shift of the plateau from the pole position is due to a large imaginary part of the residue, $A_{\rm 2^{nd}pole} \propto 3.37-i8.71$, which comes mostly from the $V_{22}(\Xi_{\rm 2^{nd}pole})$ with a strongly-positive imaginary component. At the same time, this imaginary (source) component of the $V_{22}$ is indispensable for the existence of the 2nd pole\cite{Ikeda10}. Thus, it is revealed that {\it the nature of the 2nd pole inherited from the energy-dependent $V_{22}(\Xi)$ is incompatible with the formation of a peak-shaped structure at the pole position}. Figure \ref{fig7}(b) denotes substantial contributions of the 2nd-pole term and of the remaining part when the 1st pole effect is removed in a way as explained in Fig. \ref{fig8}. The pole term and the remaining continuum part are of same order, making the strength function positive definite. This means that phenomenological assignment of only a Breit-Wigner pole-term amplitude to the 2nd pole ( for example \cite{Jido10}) is a partial procedure and is not justified.  

The 1st pole is a possible origin of the peak structure in the $\Lambda(1405)$ region, as suggested from the remaining-part curve (brown one) in Fig. \ref{fig7}(a). In fact, the residue of the 1st pole is $A_{\rm 1^{st}pole} \propto 0.01+i1.33$, the positive imaginary part of which means a peak appearance at the lower-mass side than the 1st-pole position. In order to confirm the peak formation due to the 1st pole, we introduce an arbitrary factor, $f_{11}$, to shift the 1st-pole position, as $V_{11}(E) \rightarrow f_{11}\cdot V_{11}(E)$ into $V^{\rm opt}_{22}(E)$. Figure \ref{fig8} shows the effects of the shift of the 1st-pole position on the $T_{22}$ spectrum. As the 1st pole moves towards the higher-mass side with decreasing $f_{11}$ and goes out of the $[+({\rm ch.}1),-({\rm ch.}2)]$ Riemann sheet which is relevant to the $\Lambda(1405)$, the peak structure gradually damps and disappears in spite of a persistent existence of the 2nd pole on the $[+,-]$ sheet. Thus, it is confirmed that {\it the peak structure of the $T_{22}$ spectrum comes from the 1st pole}.

\begin{figure}[hbtp]
\begin{center}
\hspace{0.5cm}
\vspace{-0.5cm}
\epsfxsize=8.5cm
\epsfbox{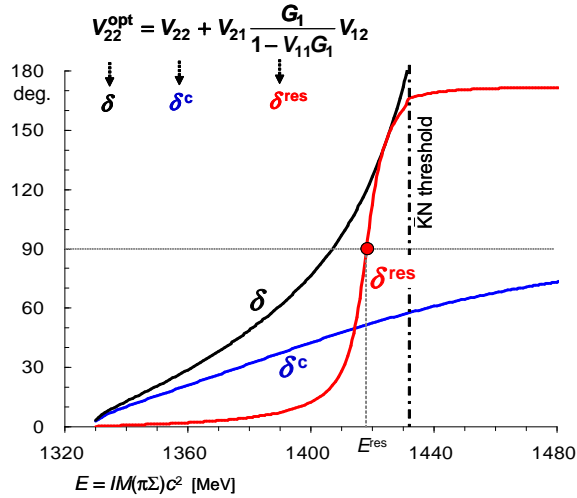}
\vspace{0.5cm}
\end{center}
\caption{Behavior of the $\pi \Sigma$ scattering phase shifts calculated from $V_{22}$, $V_{21} \frac{G_1}{1-V_{11}G_1} V_{12}$ and the total $V^{\rm opt}_{22}$.}
\label{fig9}
\end{figure}

The $T_{22}$ mass spectrum is also studied by the phase-shift contributions from each term of the optical potential of Eq. (\ref{opt4r}). Figure \ref{fig9} shows phase shifts of $\pi \Sigma$ scattering, $\delta^{\rm c}$, $\delta^{\rm res}$ and $\delta$, calculated from $V_{22}$, $V_{21} \frac{G_1}{1-V_{11}G_1} V_{12}$ and the total $V^{\rm opt}_{22}$, respectively. These phase shifts, ($\delta^{\rm c}$, $\delta^{\rm res}$ and $\delta$), are connected to the $T_{22}$ spectra of the $f_{12}$=0.0 (blue) curve of Fig. \ref{fig4}(a), the $f_{22}$=0.0 (red) curve of Fig. \ref{fig3}(a) and the $f_{12}$=1.0 (black) curve of Fig. \ref{fig4}(a), respectively, through the steps of $\delta \Rightarrow$ forward scattering amplitude $\Rightarrow$ Im $T_{22} \Rightarrow$ $T_{22}$ spectrum. The last step by the optical relation of Eq. (\ref{Spec2}) holds for real $E$ below the $\bar KN$ threshold. 
%
%%%%% Fig.9 %%%%%
Although the phase shift $\delta^{\rm c}$ of $V_{22}$ is a main concern of the 2nd pole, it does not reach $90^{\circ}$ in Fig. \ref{fig9} and the corresponding spectrum (blue curve) in Fig. \ref{fig4}(a) gives no evidence of resonance-peak formation. On the other hand, the phase shift from $V_{21} \frac{G_1}{1-V_{11}G_1} V_{12}$ and the corresponding spectrum (red curve) in Fig. \ref{fig3}(a) clearly show a resonance formation due to the 1st pole, which is determined essentially by the equation, $[1-V_{11}(\Xi)G_1(\Xi)]=0$. The coherent sum of $\delta^{\rm c}$ and $\delta^{\rm res}$ gives rise to the total phase shift, $\delta$, shown by an envelope (black curve) in Fig. \ref{fig9}.

The $T_{22}$ invariant-mass spectrum at $f_{12}$=1.0 (black curve) in Fig. \ref{fig4}(a) related to the $\delta$ can be decomposed approximately as 
\begin{equation}
\propto \left|~ e^{i\delta^{\rm c}(E)} {\rm sin}~ \delta^{\rm c}(E) + \frac {\Gamma^{\rm res}/2} { E - E^{\rm res} +i\Gamma^{\rm res}/2} \right|^2 ~k_2(E), 
\label{break}
\end{equation}
where the contribution from the $\delta^{\rm res}$ is expressed by using a Breit-Wigner amplitude with 1st-pole resonance parameters, $E^{\rm res}$ and $\Gamma^{\rm res}$. With the help of this expression the peak structure in Fig. \ref{fig4}(a) is explained as follows. The 2nd pole term plus the remaining part from $V_{22}$ provides the continuum spectrum at $f_{12}$=0.0 (blue curve). The {\it deviation} of the $T_{22}$ spectrum at $f_{12}$=1.0 (black curve) from the continuum one at $f_{12}$=0.0 (blue curve) is the {\it 1st-pole contribution} when the 1st pole is embedded in the sizable continuum. Thus, it is reconfirmed that the peak seen in the $T_{22}$ spectrum is never the original peak of the 2nd pole, but a remnant of the 1st pole after its interference with the continuum spectrum.

\section{Discussion}

In the previous sections we discussed in detail the characteristics and origin of the detectable $T_{21}$ and $T_{22}$ spectra of the coupled channel $\bar KN$-$\pi \Sigma$ meson-baryon dynamics. It is shown that the 2nd pole of $\Lambda(1405)$ is not responsible for the peak structures observed in both the $T_{21}$ and $T_{22}$ spectra. According to this finding, we will present a brief critical review on earlier theoretical analyses based on the double-pole picture of the $\Lambda(1405)$ resonance \cite{Jido10,Siebenson13,Magas05}. 

%
%%%%% Fig.10 %%%%%
\begin{figure}[hbtp]
\begin{center}
\hspace{0.5cm}
\vspace{-0.5cm}
\epsfxsize=12.5cm
\epsfbox{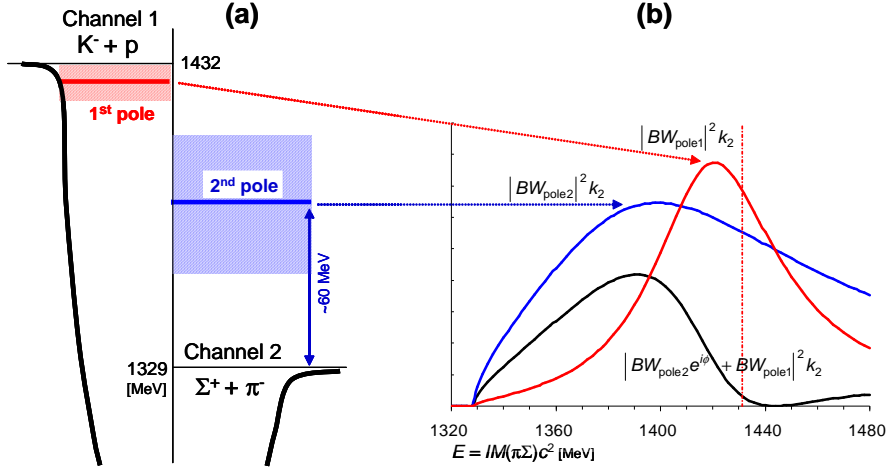}
\vspace{0.5cm}
\end{center}
\caption{(a) Schematic picture of the double poles, and (b) a spectrum model with two Breit-Wigner (B-W) amplitudes corresponding to the double poles and their coherent sum supposing the $T_{22}$ spectrum. The spectrum curves are plotted, as an example, with the best-fit parameters for HADES data given in Ref. \cite{Siebenson13}.}
\label{fig10}
\end{figure}
Figure \ref{fig10}(a) is a schematic picture of the double poles, where the 1st pole has a main component in the $\bar KN$ channel and the 2nd pole does in the $\pi \Sigma$ channel, and Fig. \ref{fig10}(b) is a model with two Breit-Wigner (B-W) amplitudes corresponding to the double poles and their coherent sum.  Jido et al. explained the behavior of the $T_{22}$ spectrum with two B-W amplitudes \cite{Jido03,Jido10}. However, we now know that their explanation is wrong because the 2nd pole term is no longer dominating, but forms a continuum background together with the sizable remaining part denoted in Fig. \ref{fig7}(b). A naive assignment of a B-W amplitude to the 2nd pole is not justified, as discussed in Subsection 3.4. 

Siebenson and Fabbietti \cite{Siebenson13} analyzed the HADES data phenomenologically by employing the two-pole model, where they assumed the $\Lambda(1405)$ consisting of double poles, and parameterized the observed peak as a coherent sum of the B-W amplitudes. Although the HADES data was well reproduced with double poles of ($M_{\rm pole}c^2, \Gamma$)=(1418, 58) and (1375, 146) MeV, there is no physical reason to relate one of the B-W amplitudes to the chiral 2nd pole, because the 2nd pole produces no peak structure in the detectable spectrum. On the other hand, Hassanvand et al.\cite{Hassanvand13} treated the same HADES data in a framework of single pole, and determined a best-fitted value of ($M_{\rm pole}c^2, \Gamma$)=($1405^{+11}_{-9}, 62 \pm10$) MeV with statistical confidence intervals. This single-pole fit is more successful, contributing to the present PDG value of the $\Lambda(1405)$, 1405.1 MeV, than the above-mentioned double-pole fit.

One evidence for the two-pole nature of the $\Lambda(1405)$ was claimed by Magas {\it et al.}\cite{Magas05} through their analyses on reactions, $K^- p \rightarrow \pi^0 \pi^0 \Sigma^0$ of Prakhov {\it et al.}\cite{Prakhov04} and $\pi^-p \rightarrow K^0 \pi \Sigma$ of Thomas {\it et al.}\cite{Thomas73} It was argued that the former reaction has a higher probability of forming an intermediate state of the 1st pole (in that reference, the second state with higher mass), while the latter reaction favors the formation of the 2nd pole (the first state with lower mass). Then, the different shapes of these two reaction spectra were interpreted as being evidence of the two poles. But there is no connection between the 2nd pole and the peak behavior of Thomas {\it et al.}'s data, since neither of the basic $T_{22}$ and $T_{21}$ spectra has peaks related to the 2nd pole, as we discussed in Section 3. In spite of the authors' claim, no evidence for the chiral double poles of the $\Lambda(1405)$ was obtained there. Recently, some results of the chiral double poles have been listed in Meissner and Hyodo's table \cite{Meissner15}. It should be mentioned, concerning the findings given in Subsections 3.3 and 3.4, that the listed 2nd pole positions and widths are of supplementary information on not peak but continuum contributions to the spectrum for the $\Lambda(1405)$, as long as they are concerned with the observed CLAS data.

\section{Conclusion}

In this paper we have given detailed accounts of the characteristics and origin of detectable $T_{21}$ and $T_{22}$ spectra in the $\bar KN$-$\pi \Sigma$ coupled channel system. With the use of generalized optical potentials, each spectrum is decomposed into contributions from the 1st pole, the 2nd pole and the remaining part. Summary results are given in Figs. \ref{fig2}(a)(b) and \ref{fig8}. The 1st pole is responsible for both the peak structures in the $T_{21}$ and $T_{22}$ invariant-mass spectra. The 2nd pole due to the energy-dependent chiral interaction provides a continuum amplitude appreciably affecting the shape of the peak formed by the 1st pole in the detectable $T_{22}$ spectrum. In other words, the supposition of a peak structure due to the 2nd pole around the 2nd pole position is of total misunderstanding, as long as it is concerned with the experimentally observed $\Lambda(1405)$. 

Finally, it is concluded that experimentally observed $\Lambda (1405)$ resonance is not of double-pole nature, but {\it of single-pole nature} formed as the Feshbach resonance \cite{Feshbach58}. No evidence for the chiral 2nd pole can be obtained from any experimental data in the relevant $\Lambda(1405)$ mass region, because all detected decay particles are on-shell. The single-pole nature of the $\Lambda(1405)$ would provide a unique starting basis\cite{Akaishi02,Yamazaki02} for current studies of the production and the structure of anti-kaonic nuclei.

\section*{Acknowledgments}

The authors thank Prof. O. Morimatsu and Prof. K. Yazaki for critical discussions. They acknowledge the receipt of Grant-in-Aid for Scientific Research of Monbu-Kagakusho of Japan.

\end{document}